# Linac*LHC Based ep, γp, eA, γA and FELγ-A Colliders: Luminosity and Physics

A.K. Çiftçi[a], S. Sultansoy[b,c], Ö. Yavaş[d]

[a]Dept. of Physics, Faculty of Science, Ankara University, 06100 Tandogan, Ankara, TURKEY
[b]DESY, Notke Str. 85, D-22607 Hamburg, GERMANY,
[c]Institute of Physics, Academy of Sciences, H. Cavid Ave. 33, Baku, AZERBAIJAN
[d]Dept. of Eng. Physics, Faculty of Science, Ankara University, 06100 Tandogan, Ankara, TURKEY

## Abstract

Main parameters and physics goals of different colliders, which can be realized if a special 1 TeV energy linear electron accelerator or corresponding linear collider is constructed tangential to LHC, are discussed. It is shown that $L_{ep}=10^{32}$ cm$^{-2}$s$^{-1}$ at $\sqrt{s_{ep}}$=5.29 TeV can be achieved within moderate upgrade of LHC parameters. Then, γp collider with the same luminosity and $\sqrt{s}$=4.82 TeV can be realized using Compton backscattering of laser beam off the electron beam. Concerning the nucleus beam, $L*A=10^{31}$ cm$^{-2}$s$^{-1}$ can be achieved at least for light and medium nuclei both for eA and γA options. Finally, colliding of the FEL beam from an electron linac with nucleus beams from LHC will give a new opportunity to investigate nuclear spectroscopy and photo-nuclei reactions.

PACS Numbers: 13.60.-r , 24.85.+p

Keywords: Collider, LHC, Linac, Photon, Hadron, Nucleus

## 1 Introduction

It is well known that lepton-hadron collisions play a crucial role in our understanding of micro-world. An excellent example is the discovery of the quark-parton structure of nucleons. At the fixed target experiments the region $Q^2$<1000 GeV$^2$ is explored. The HERA collider enlarged this region by two orders. The possible LEP*LHC collider has an important disadvantage of $E_e/E_p$<0.01. The synchrotron radiation restricts the electron energy obtainable at ring machines and a transition to linear accelerators seems unavoidable for $E_e$>100 GeV. Therefore, one should consider linac-ring type machines in order to achieve TeV scale at constituent level in lepton-hadron collisions. The possible alternative, namely μp collider will face even more problems than the basic $\mu^+\mu^-$ collider.

In this paper we consider collisions of electron and photon beams obtained from 1 TeV linear accelerator with proton and nucleus beams from the LHC. The center of mass energies which will be achieved at different options of this machine are an order larger than those at HERA are and ~3 times larger than the energy region of TESLA⊗HERA, LEP⊗LHC and $\mu$-ring⊗TEVATRON (see the review [1]). Following [1-4] below we consider electron linac with $P_e \approx 60$ MW (Table 1) and upgraded proton beam from LHC (Table 2). The reasons for choosing superconducting linac, instead of conventional warm linacs (NLC, JLC) or CLIC, are listed in [2].

**2 Main parameters and physics search potential of ep collider**

According to Tables 1 and 2, center of mass energy and luminosity for this option are $\sqrt{s}$=5.29 TeV and $L_{ep}=10^{32}$cm$^{-2}$s$^{-1}$, respectively, and an additional factor 3-4 can be provided by the "dynamic" focusing scheme [5]. Further increasing will require cooling at injector stages. This machine, which will extend both the $Q^2$–range and $x$-range by more than two order of magnitude comparing to those explored by HERA, has a strong potential for both SM and BSM research. For example: the discovery limit for the first generation leptoquarks is m ≈ 3 TeV; the discovery limit for SUSY particles is $m_{\tilde{l}} + m_{\tilde{q}} \approx 1.5$ TeV and covers all six SUGRA points, which are used for SUSY analyses at LHC [6]; excited electron will be copiously produced up to $m_{e*} \approx 2$ TeV etc.

**3 Main parameters and physics search potential of γp collider**

The advantage in spectrum of back-scattered photons (see [7] and references therein) and sufficiently high luminosity (for details see ref. [8,9]), $L_{\gamma p}>10^{32}$cm$^{-2}$s$^{-1}$ at $z$=0, will clearly manifest itself in a searching of different phenomena. In Fig. 1 the dependence of luminosity on the distance $z$ between interaction point (IP) and conversion region (CR) is plotted (for corresponding formulae see [9]). In Fig. 2 we plot luminosity distribution as a function of γp invariant mass $W_{\gamma p}=2\sqrt{E_\gamma E_p}$ at $z=5$ m. In Fig. 3 this distribution is given for the choice of $\lambda_e=0.8$ and $\lambda_0=-1$ at three different values of the distance between IP and CR.

The physics search potential of γp colliders is reviewed in [10]. The γp option will essentially enlarge the capacity of the Linac*LHC complex. For example, thousands di-jets with $p_t$>500GeV and hundreds thousands single W bosons will be produced, hundred millions of $b^*b$ and $c^*c$ pairs will give opportunity to explore the region of extremely small $x_g$ (~$10^{-6}$-$10^{-7}$) etc. Concerning the BSM physics:

- linac-ring type γp colliders are ideal machines for $u^*, d^*$ and $Z_8$ search (the discovery limits are $m_{u^*}=4$ TeV, $m_{d^*}=3$ TeV and $m_{Z_8}=3$ TeV)

- the fourth SM family quarks (the discovery limits are $m_{u_4}\approx 0.8$ TeV and $m_{d_4}\approx 0.7$ TeV) will be copiously produced, since their masses are predicted to be in the region 300÷700 GeV with preferable value $m_{u_4}\approx m_{d_4}\approx 640$ GeV (see [11] and references therein)

- because of parameter inflation, namely more than 160 observable free parameters in the three family MSSM (see [12] and references therein), SUSY should be realized at preonic level. Nevertheless, the discovery limits for different channels are $m_{\tilde{W}}+m_{\tilde{q}}\approx 1.4$ TeV, $m_{\tilde{g}}+m_{\tilde{q}}\approx 1.2$ TeV, $m_{\tilde{\gamma}}+m_{\tilde{q}}\approx 0.4$ TeV and $m_{\tilde{q}}+m_{\tilde{q}}\approx 1.2$ TeV.

**4 Main parameters and physics search potential of eA collider**

In the case of LHC nucleus beam IBS effects in main ring are not crucial because of large value of $\gamma_A$. The main principal limitation for heavy nuclei coming from beam-beam tune shift may be weakened using flat beams at collision point. Rough estimations show that $L_{eA}\cdot A>10^{31}$cm$^{-2}$s$^{-1}$ can be achieved at least for light and medium nuclei [1, 13]. By use of parameters of nucleus beams given in Table 3 one has $L_{eC}\cdot A=10^{31}$cm$^{-2}$s$^{-1}$ and $L_{ePb}\cdot A=1.2\cdot 10^{30}$cm$^{-2}$s$^{-1}$, correspondingly. This machine will extend both the $Q^2$–range and $x$-range by more than two orders of magnitude with respect to the region, which can be explored by HERA based eA collider [14].

## 5 Main parameters and physics search potential of γA collider

Limitation on luminosity due to beam-beam tune shift is removed in the scheme with deflection of electron beam after conversion [9]. The dependence of luminosity on the distance between IP and CR for γC and γPb options are plotted in Figs. 4 and 5, respectively. As it is seen from the plots, $L_{\gamma C} \cdot A = 0.8 \cdot 10^{31} cm^{-2} s^{-1}$ and $L_{\gamma Pb} \cdot A = 10^{30} cm^{-2} s^{-1}$ at $z=5$ m. Center of mass energy of Linac*LHC based γA collider corresponds to $E_\gamma \sim$ PeV in the lab system. At this energy range cosmic ray experiments have a few events per year, whereas γ-nucleus collider will give few billions events. This machine has a wide research capacity. The list of physics goals contains:

- total cross sections to clarify real mechanism of very high energy γ - nucleus interactions
- investigation of hadronic structure of photon in nuclear medium
- according to VMD, proposed machine will be also $\rho$ - nucleus collider
- formation of the quark-gluon plasma at very high temperatures but relatively low nuclear densities
- gluon distribution at extremely small $x_g$ in nuclear medium ( $\gamma A \to Q\bar{Q} + X$ )
- investigation of both heavy quark and nuclear medium properties ( $\gamma A \to J/\Psi + X$, $J/\Psi \to l^+ l^-$ )
- etc.

Especially, the investigation of gluon distribution at extremely small $x_g$ will give a crucial information for QCD in nuclear medium.

## 6 FELγ–A Collider

The ultra-relativistic ions will see laser photons with energy $\omega_0$ as a beam of photons with energy $2\gamma_A \omega_0$, where $\gamma_A$ is the Lorentz factor of the ion beam. For LHC $\gamma_A = (Z/A)\gamma_p = 7446(Z/A)$, therefore, $0.1 \div 10$ keV photons, produced by the linac based FEL, correspond to $0.5 \div 50$ MeV in the nucleus rest frame. The huge number of events and small energy spread of colliding beams [15] will give opportunity to scan an interesting region with adjusting of FEL energy.

## 7 Conclusion

The proposed complex, if realised, will open new horizons for both the particle and the nuclear physics. Therefore, it is necessary to continue the efforts on both machine and physics search potential aspects.

**Acknowledgements**

We are grateful to R. Brinkmann, M. Leenen, D. Trines, G.-A. Voss and F. Willeke for useful discussion and valuable remarks. S. Sultansoy is grateful to DESY Directorate for invitation and hospitality. This work is supported by Turkish State Planning Organization under the Grant No DPT-97K-120420 and DESY.

Table 1. Parameters of special electron linac

| | |
|---|---|
| Electron energy, GeV | 1000 |
| No of electrons per bunch, $10^{10}$ | 0.7 |
| Bunch length, mm | 1 |
| Bunch spacing, ns | 100 |
| No of bunches per pulse | 5000 |
| Pulse Length, μs | 1000 |
| Repetition rate, Hz | 10 |
| Beam power, MW | 56 |
| Normalised emittance, $10^{-6}$m | 10 |
| Beta function at IP, cm | 200 |
| $\sigma_{x,y}$ at IP, μm | 3.3 |
| Beta function at CR, cm | 2 |
| $\sigma_{x,y}$ at CR, μm | 0.33 |

Table 2. Upgraded parameters of LHC proton beam

| Proton energy, GeV | 7000 |
|---|---|
| No of protons per bunch, $10^{10}$ | 40 |
| Bunch spacing, ns | 100 |
| Normalised emittance, $10^{-6}$m | 0.8 |
| Bunch length, cm | 7.5 |
| Beta function at IP, cm | 10 |
| $\sigma_{x,y}$ at IP, µm | 3.3 |

Table 3. Parameters of C and Pb beams

| | C | Pb |
|---|---|---|
| Nucleus energy, TeV | 42 | 574 |
| Particles per bunch, $10^{10}$ | 1 | 0.01 |
| Normalised emittance, $10^{-6}$m | 1.25 | 1.4 |
| Bunch length, cm | 7.5 | 7.5 |
| Beta function at IP, cm | 10 | 10 |
| $\sigma_{x,y}$ at IP, µm | 5.8 | 6.9 |
| Bunch spacing, ns | 100 | 100 |

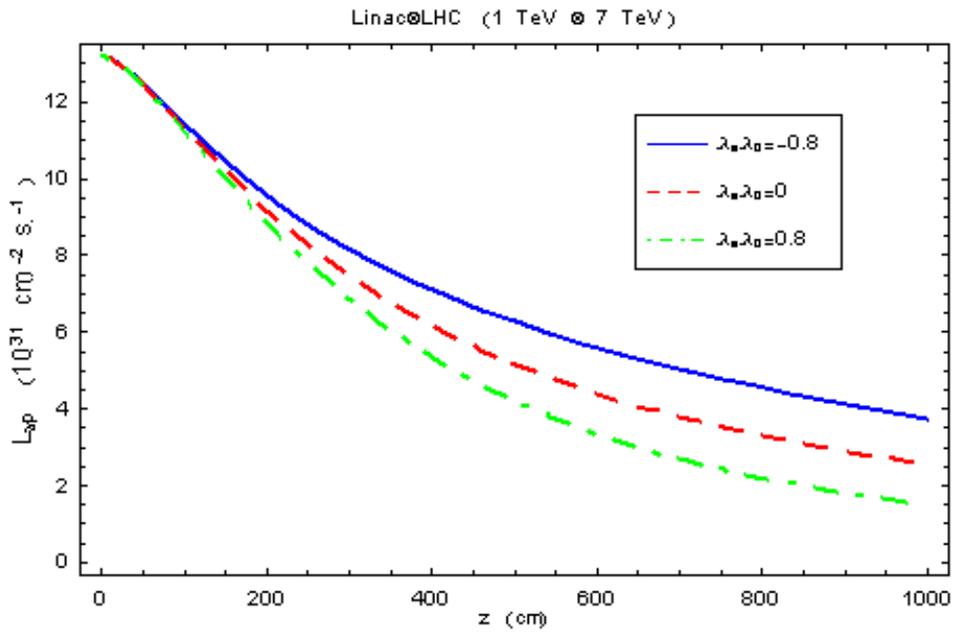

Figure 1. The dependence of luminosity on the distance z for γp collider.

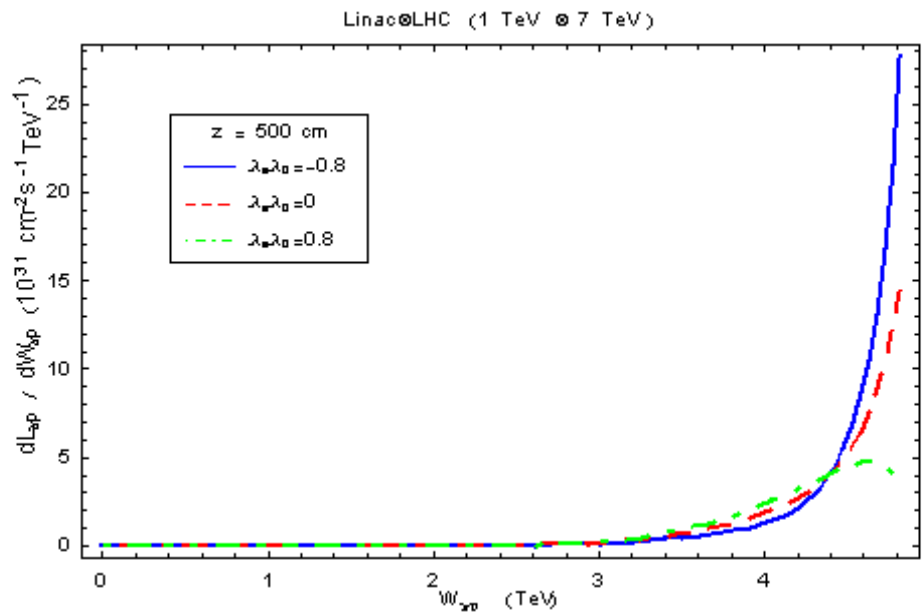

Figure 2. Luminosity distribution as a function of γp invariant mass ($W_{\gamma p}$) at z=5m for choice of three different electron polarization.

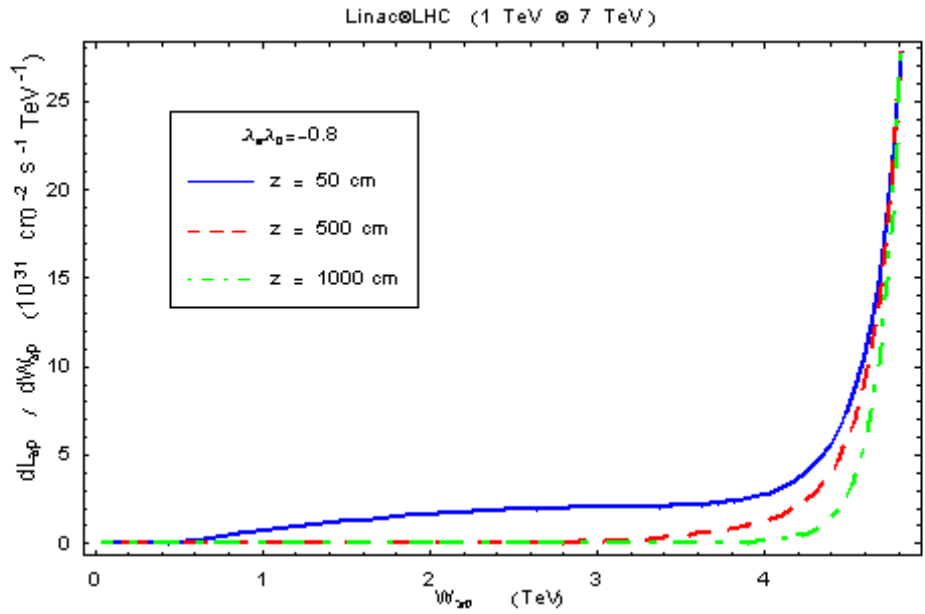

Figure 3. Luminosity distribution as a function of γp invariant mass ($W_{\gamma p}$) for three different z values.

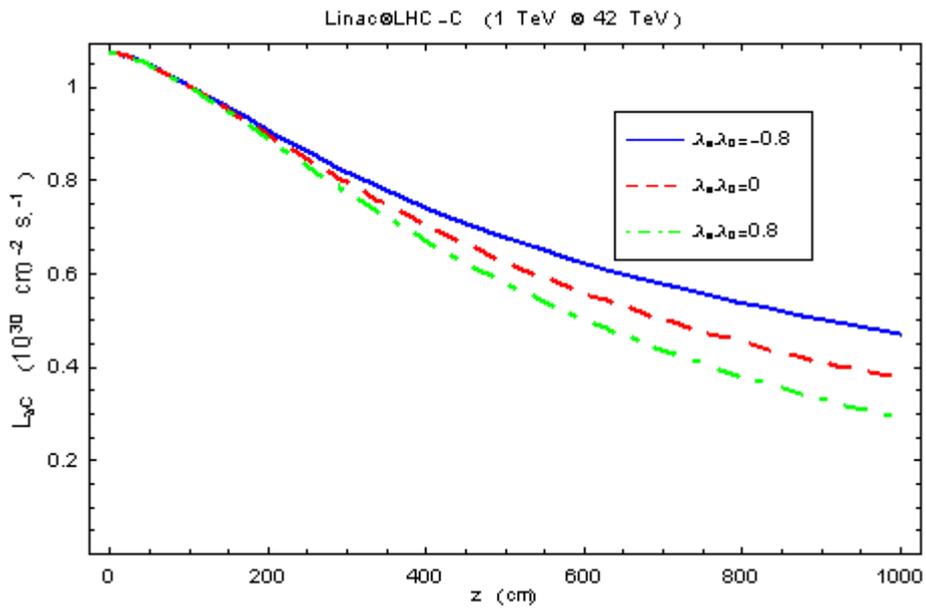

Figure 4. The dependence of luminosity on the distance for γC collider

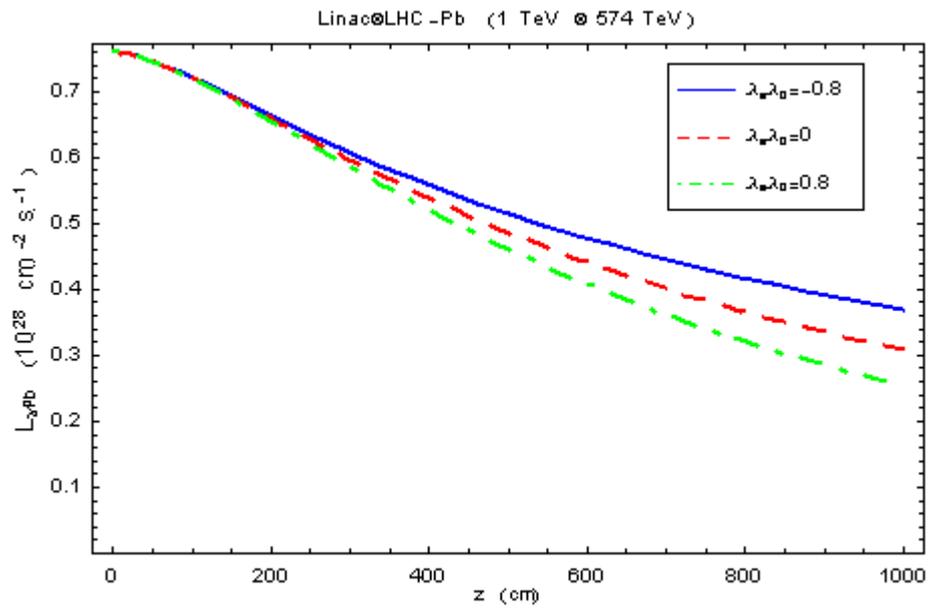

Figure 5. The dependence of the luminosity on the distance z for γPb collider.